\documentclass[twocolumn,showpacs,preprintnumbers,superscriptaddress]{revtex4}
\usepackage{amssymb}
\usepackage{amsmath}
\usepackage{dcolumn}
\usepackage{bm}
\usepackage{graphicx}
\usepackage{subfigure}

\begin{document}
%\begin{CJK*}{GBK}{song}

\title{Dynamics of Weyl quasiparticles emerged in an optical lattice}
\author{Zhi Li}
\affiliation{National Laboratory of Solid State Microstructures
and School of Physics, Nanjing University, Nanjing 210093,
China}\affiliation{Guangdong Provincial Key Laboratory of Quantum
Engineering and Quantum Materials, SPTE, South China Normal
University, Guangzhou 510006, China}

\author{Huai-Qiang Wang}
\affiliation{National Laboratory of Solid State Microstructures
and School of Physics, Nanjing University, Nanjing 210093, China}

\author{Dan-Wei Zhang}
\affiliation{Guangdong Provincial Key Laboratory of Quantum
Engineering and Quantum Materials, SPTE, South China Normal
University, Guangzhou 510006, China}

\author{Shi-Liang Zhu}
\email{slzhu@nju.edu.cn} \affiliation{National Laboratory of Solid
State Microstructures and School of Physics, Nanjing University,
Nanjing 210093, China} \affiliation{Synergetic Innovation Center
of Quantum Information and Quantum Physics, University of Science
and Technology of China, Hefei, Anhui 230026, China}

\author{Ding-Yu Xing}
\email{dyxing@nju.edu.cn} \affiliation{National Laboratory of
Solid State Microstructures and School of Physics, Nanjing
University, Nanjing 210093, China} \affiliation{Collaborative
Innovation Center of Advanced Microstructures, Nanjing University,
Nanjing 210093, China}
\date{\today}

\begin{abstract}
We investigate the dynamics of the Weyl quasiparticles emerged in
an optical lattice where the topological Weyl semimental and
trivial band insulator phases can be adjusted with the on-site
energy. The evolution of the density distribution is demonstrated
to have an anomalous velocity in Weyl semimental but a steady
\emph{Zitterbewegung} effect in the band insulator. Our analysis
demonstrates that
%the topological Chern number and
the chirality
of the system can be directly determined from the positions of the
atomic center-of-mass. Furthermore, the amplitude and the period
of the relativistic \emph{Zitterbewegung} oscillations are shown
to be observable with the time-of-flight experiments.
\end{abstract}
\pacs{67.85.-d, 03.65.Vf, 03.75.Kk} \maketitle

\section{Introduction}
Weyl equation, first proposed by Hermann Weyl, is a relativistic
wave equation to describe massless spin-1/2 particles in quantum
field theory \cite{Weyl}. But such fermions (so-called Weyl
fermions) have not been observed as fundamental particles in
nature \cite{Balents}. Recently, it was demonstrated that a Weyl
fermion can emerge as a quasiparticle in condensed matter
\cite{Wan,Xu1,Burkov} or photonic crystals \cite{Lu,Lu2014}-Weyl
semimetal (WSM). In WSM, two linear dispersion bands in
three-dimensional (3D) momentum space intersect as a single
degenerate point-the Weyl point, which is a monopole of Berry flux
with topological charge defined by the Chern number. Many
intriguing features of the WSM, such as the topologically
protected Fermi arc on the surface and the Weyl points in the
bulk, have been experimentally observed in condensed matter
\cite{Xu2015,Lv} or photonic crystals \cite{Lu2015}; however, the
fundamental dynamics of the Weyl equation is hard to detect in
such systems. On the other hand, it was recently proposed that the
WSM can be realized with ultracold atoms loaded into a tunable
cubic lattice \cite{Zhang1,Dubcek,Xu,He}. This provides a
versatile platform to study the dynamics of the Weyl
quasiparticles, which has not yet been investigated.

In this paper, we exploit the dynamics of the Weyl quasiparticles
emerged in an optical lattice, which is otherwise difficult to do
in condensed matter systems. We consider a Gaussian wave packet
which is formed by a harmonic trap and initially placed at a fixed
Weyl point in the optical lattice. After releasing the trap, the
atoms would start expanding in the 3D Weyl lattice, and hence the
dynamics of the atoms, which is described by the Weyl equation,
can be exploited with the time-of-flight (TOF) experiments
(hereafter the ``TOF" denotes the evolution experiments of atomic
gases in the Weyl lattice rather than the traditional TOF where
the lattice potential is also released.). We calculate the
time-dependent density profiles and find that the 3D Gaussian wave
packet in WSM gradually evolves into a two-layer spherical shell
structure because of the interference of the positive and negative
energy states. We then demonstrate that the topological Chern
number and the chirality of the system can be directly determined
from the time-dependent position of the center-of-mass (PCM).
Furthermore, we show that the amplitude and the period of the
relativistic \emph{Zitterbewegung} (ZB) in the trivial band
insulator (BI) are sufficiently large to be observable. Notably,
the topological Chern number and ZB have not yet been directly
observed in condensed matter systems. Since the density profiles
can be readily observed with TOF experiments, our work may
stimulate the experimental study of the dynamics of the Weyl
fermions in optical lattices.

The paper is organized as follows. In Sec. II, we review the
scheme proposed in Ref.\cite{Dubcek} for realizing the Weyl
quasiparticle with cold atomic gases in optical lattice. In Sec.
III, we investigate and discuss the dynamics of Weyl
quasiparticles in 3D optical lattices. We give a brief summary in
section IV.  The split-operator method used in this work and some
computational details for topological invariants are listed in
Appendix.

\section{model}
The model under consideration is a cubic lattice with phase
engineered hopping along $x$- and $z$-directions, which possesses
Weyl points in the momentum space. The tight-binding Hamiltonian
is given by the form \cite{Dubcek}
\begin{equation}\label{H discrete}
\begin{split}
&\hat{H}=-\sum_{m,n,l}\{[J_xe^{-i\Phi_{m,n,l}}\hat{a}^{\dag}_{m+1,n,l}
+J_{z}e^{-i\Phi_{m,n,l}}\hat{a}^{\dag}_{m,n,l+1}\\&+J_{y}\hat{a}^{\dag}_{m,n+1,l}+(-1)^{(m+n-1)}\frac{\varepsilon}{2}
\hat{a}^{\dag}_{m,n,l}]\hat{a}_{m,n,l}+\textrm{H.c.}\},
\end{split}
\end{equation}
where $J_{x,y,z}$ denote the tunnelling amplitudes,
$\hat{a}^{\dag}_{m,n,l}$ ($\hat{a}_{m,n,l}$) is the creation
(annihilation) operator on the site ($m,n,l$), and
$\Phi_{m,n,l}=m\Phi_x+n\Phi_y+l\Phi_z$ are the nontrivial hopping
phases. As shown in Fig. 1(a)(b), the lattice is a stacking of 2D
Harper-Hofstadter lattice which has already been realized
experimentally \cite{Aidelsburger1,Aidelsburger2,Miyake}, and the
$z$-direction hopping has phase $0$ ($\pi$) for $m+n$ even (odd).
The lattice has two sublattices (A-B) giving rise to pseudospin.
The on-site energy is $\varepsilon/2$ ($-\varepsilon/2$) for $m+n$
odd (even), which is the key parameter for manipulating the
topological phase transition of the system. The positions of the
sites can be expressed as $\mathbf{R}_{m,n,l}
=m\mathbf{a_1}+n\mathbf{a_2}+l\mathbf{a_3}$, where $m$, $n$ and
$l$ are integers, and $\mathbf{a_1}=(a,0,0)$,
$\mathbf{a_2}=(0,a,0)$ and $\mathbf{a_3}=(0,0,a)$ (hereafter we
set the lattice spacing $a=1$). By introducing Fourier transform
$\hat{a}_{m,n,l}^{\dag}=\sum_{\mathbf{k}}e^{i\mathbf{k}\cdot\mathbf{R}_{m,n,l}}\hat{a}_\mathbf{k}^{\dag}$,
where $\mathbf{k}=(k_x,k_y,k_z)$ is the Bloch wave vector, we
obtain the Hamiltonian in quasimomentum representation given by
\begin{equation}\label{Hk}
\begin{split}
H_{k}=&-2J_{y}\cos{k_{y}}\sigma_{x}-2J_{x}\sin{k_{x}}\sigma_{y}+(\varepsilon+2J_{z}\cos{k_{z}})\sigma_{z},
\end{split}
\end{equation}
where the Pauli matrices $\sigma_j\ (j=x,y,z) $ are pseudospins
describing A and B sublattices of the system. The WSM is usually
realized when the time-reversal symmetry or inversion symmetry is
broken. The \textbf{k}-space Hamiltonian in Eq. (\ref{Hk}) has
time reversal symmetry, $H(\mathbf{k})^*=H(-\mathbf{k})$, but
inversion symmetry is broken here because $\sigma_x
H(\mathbf{k})\sigma_x\neq H(-\mathbf{k})$. Nevertheless, since the
quasiparticles' evolution we obtained is entirely derived from
the low-energy effective Hamiltonian and depends totally on the
structure and topological property of Weyl points, these results
are therefore also valid for the case of WSM with broken time
reversal symmetry which has the same spectrum. The energy spectrum
of the Bloch bands is given by
\begin{equation}\label{Ek}
\begin{split}
E_\mathbf{k}=\pm\sqrt{4J_x^2\sin^2k_x+4J_y^2\cos^2{k_y}+(\varepsilon+2J_z\cos{
k_z})^2},
\end{split}
\end{equation}
which is plotted in Figs. 1(c-e) for typical values of
$\varepsilon$.

%%%%%%%%%%%%%%%%%%%%%%%%%%%%%%%%%%%%%%%%%%%%%%%%%%%%%%%%%%%%%%%%%%%%%%%%%%%%%%%%%%%%%%%%%%%%%%%%%%%%%%%%%%%%%%%%%%%%%%%%%%%%%%
%%%%%%%%%%%%%%%%%%%%%%%%%%%%%%%%%%%%%%%%%%%%%%%%%%%%%%%%%%%%%%%%%%%%%%%%%%%%%%%%%%%%%%%%%%%%%%%%%%%%%%%%%%%%%%%%%%%%%%%%%%%%%%
\begin{figure}[htbp] \centering
\label{Fig1}
\includegraphics[width=8cm]{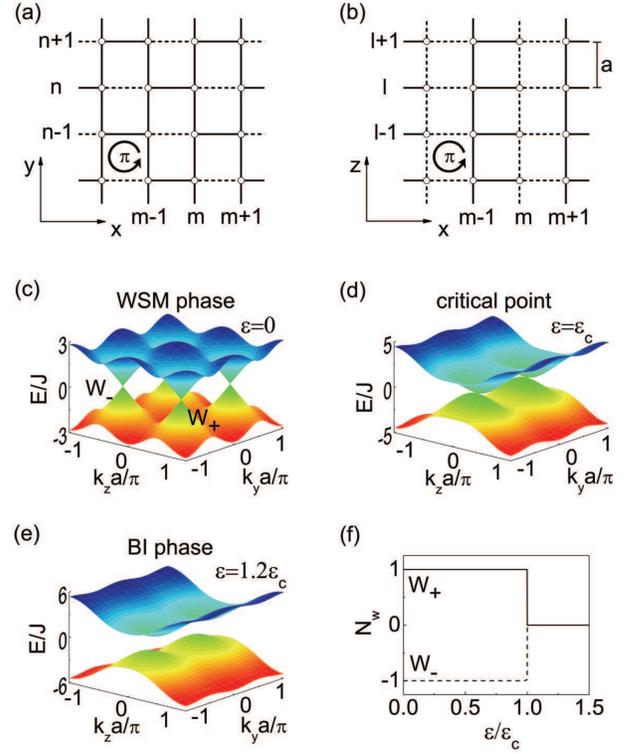}
 \caption{(Color online).
(a,b) Schematic view of the tight-binding model described in Eq.
(\ref{H discrete}). Solid and dash lines depict hopping with
acquired phase 0 and $\pi$, respectively. The dispersion relations
in the $k_{x}=0$ cross-section for (c) WSM phase with
$\varepsilon=0$, (d) critical point $\varepsilon=\varepsilon_{c}$
and (e) BI phase with $\varepsilon=1.2\varepsilon_{c}$. (f) The
winding number $N_{w}$ as a function of $\varepsilon$. }
\end{figure}
%%%%%%%%%%%%%%%%%%%%%%%%%%%%%%%%%%%%%%%%%%%%%%%%%%%%%%%%%%%%%%%%%%%%%%%%%%%%%%%%%%%%%%%%%%%%%%%%%%%%%%%%%%%%%%%%%%%%%%%%%%%%%%
%%%%%%%%%%%%%%%%%%%%%%%%%%%%%%%%%%%%%%%%%%%%%%%%%%%%%%%%%%%%%%%%%%%%%%%%%%%%%%%%%%%%%%%%%%%%%%%%%%%%%%%%%%%%%%%%%%%%%%%%%%%%%%

The system is a WSM phase for $\varepsilon<|\varepsilon_c|$ and a
band insulator for $\varepsilon>|\varepsilon_c|$ with the critical
points $\varepsilon_c=\pm 2J_z$. In WSM phase, the energy spectra
touch at four Weyl points (two pairs) in the first Brillouin zone
at $\mathbf{W}_{\pm}=(0, \pm\frac{\pi}{2},
\pm\arccos{\frac{-\varepsilon}{2J_z})}$. By making $\varepsilon$
large enough ($\varepsilon>|\varepsilon_c|$), the Weyl points with
opposite chiralities can be driven to annihilate at
$\mathbf{M}=(0,\pm\frac{\pi}{2},0)$ for $\varepsilon_c=-2J_{z}$ or
at the edge of BZ for $\varepsilon_c=2J_{z}$. Without loss of
generality, we only consider the case of $\varepsilon_c=-2J_{z}$,
and set $\mathbf{k}=\mathbf{W}_{\pm}+\mathbf{q}$, where
$\mathbf{q}=(q_{x},q_{y},q_{z})$ is the displacement vector
measured from the Weyl point $\mathbf{W}_{\pm}$ in momentum space.
By expanding the quasi-momentum \textbf{k} at $\mathbf{W}_{\pm}$,
the low-energy effective spectrum can be obtained as
\begin{equation}\label{E2}
\begin{aligned}
E_{\mathbf{q}}&=\pm
\sqrt{\alpha_{x}^{2}{v}_{x}^2q_{x}^2+\alpha_{y}^{2}{v}_{y}^2q_{y}^2+(\Delta+\alpha_{z}
{v}_{z}q_{z}+\frac{\hbar^2 q_{z}^2}{2{m}^*})^2}
\end{aligned}
\end{equation}
with $\alpha_{x,y,z}=\pm 1$ and
$v_{x,y,z}=(2J_{x},2J_{y},\sqrt{4J_z^2-\varepsilon^2})/\hbar$,
$\Delta=0$ and ${m}^*=1/\varepsilon$ for the WSM phase;
${v}_{x,y,z}=(2J_{x},2J_{y},0)/\hbar$,
$\hbar\Delta=\varepsilon-\varepsilon_{c}$ and
${m}^*=1/\varepsilon_{c}$ for the BI phase. For simplicity,
throughout we set $J_{x,y,z}=J=1$ as the energy unit. In the
regime of WSM phase far away from the critical point
$\varepsilon_c$, the quadratic term can be neglected. Then
$E_{\mathbf{q}}$ exhibits a typical Weyl point spectrum
$E_{\mathbf{q}}=\pm
\sqrt{{v}_{x}^{2}q_{x}^2+{v}_{y}^{2}q_{y}^2+{v}_{z}^2q_{z}^2}$.
The Fermi velocities in the three orthogonal directions are equal
for $\varepsilon=0$, so we denote it as ${v}_{x,y,z}= v_{F}$.
However, when the system is very close to the phase transition
point, the value of ${v}_{z}$ tends to zero, the ${v}_{z}q_z$ and
the $\hbar^2 q_z^2/(2{m}^*)$ terms in Eq. (\ref{E2}) are of the
same order. The quadratic term cannot be neglected. At the
critical point, as ${v}_{z}=0$ exactly, the linear term disappears
and the quadratic term becomes dominant. This causes the hybrid
spectrum $E_{\mathbf{q}}=\pm
\sqrt{{v}_{x}^{2}q_{x}^2+{v}_{y}^{2}q_{y}^2+(\Delta+\frac{\hbar^2
q_z^2}{2{m}^*})^2}$, which is linear in $x$- and $y$-direction but
quadratic in $z$-direction.
%After the transition, a gap opens
%between the hybrid bands.

The wave function for the low-energy quasiparticles around the
Weyl point satisfies the equation of motion
\begin{equation}\label{ME}
i\hbar\partial_{t}\Psi=\hat{H}_{\textrm{eff}}\Psi,
\end{equation}
where the effective Hamiltonian
\begin{equation}\label{g}
H_{\textrm{eff}}=\vec{\sigma}\cdot\vec{g}
\end{equation}
with
\begin{equation}\label{gg}
\vec{g}=(\alpha_{y}v_{y}{q}_{y},\alpha_{x}v_{x}
{q}_{x},\Delta+\alpha_{z}v_{z}{q}_{z}+\frac{\hbar^2
{q}_{z}^2}{2m^*}).
\end{equation}

It is a 3D relativistic Hamiltonian which is valid for describing
the dynamics of the system in the whole process of the phase
transition from a Weyl semimetallic to a band insulating phase.
Notably, the 2D atomic Dirac fermions and the related topological
phase transition \cite{Zhu,Hou,Goldman,Zhang2012,
Montambaux1,Montambaux2} have been experimentally observed by
several groups \cite{Tarruell,Gomes,Duca}. The system can be
characterized by the winding number defined by \cite{Volovik}
\begin{equation}\label{Nw}
\begin{aligned}
N_\text{w}=\frac{1}{8\pi}\epsilon_{ijk}\int_{\Sigma_{2}}dS^{k}\hat{g}\cdot(\frac{\partial\hat{g}}{\partial
q_{i}}\times \frac{\partial\hat{g}}{\partial q_{j}}),
\end{aligned}
\end{equation}
where the unit vector $\hat{g}=\vec{g}/|\vec{g}|$, and $dS^k$ is
an area element of the sphere $\Sigma_{2}$ around the singular
point in the momentum space. The winding number $N_\text{w}$
versus $\varepsilon$ for Weyl points with opposite chiralities are
plotted in Fig. 1(f). It is clear that
 $N_{w}=\pm 1$ in the WSM phase and zero in the
BI phase.

%%%%%%%%%%%%%%%%%%%%%%%%%%%%%%%%%%%%%%%%%%%%%%%%%%%%%%%%%%%%%%%%%%%%%%%%%%%%%%%%%%%%%%%%%%%%%%%%%%%%%%%%%%%%%%%%%%%%%%%%%%%%%%
%%%%%%%%%%%%%%%%%%%%%%%%%%%%%%%%%%%%%%%%%%%%%%%%%%%%%%%%%%%%%%%%%%%%%%%%%%%%%%%%%%%%%%%%%%%%%%%%%%%%%%%%%%%%%%%%%%%%%%%%%%%%%%
\begin{figure*}[htbp] \centering
 %Requires \usepackage{graphicx}
\includegraphics[width=0.95\linewidth]{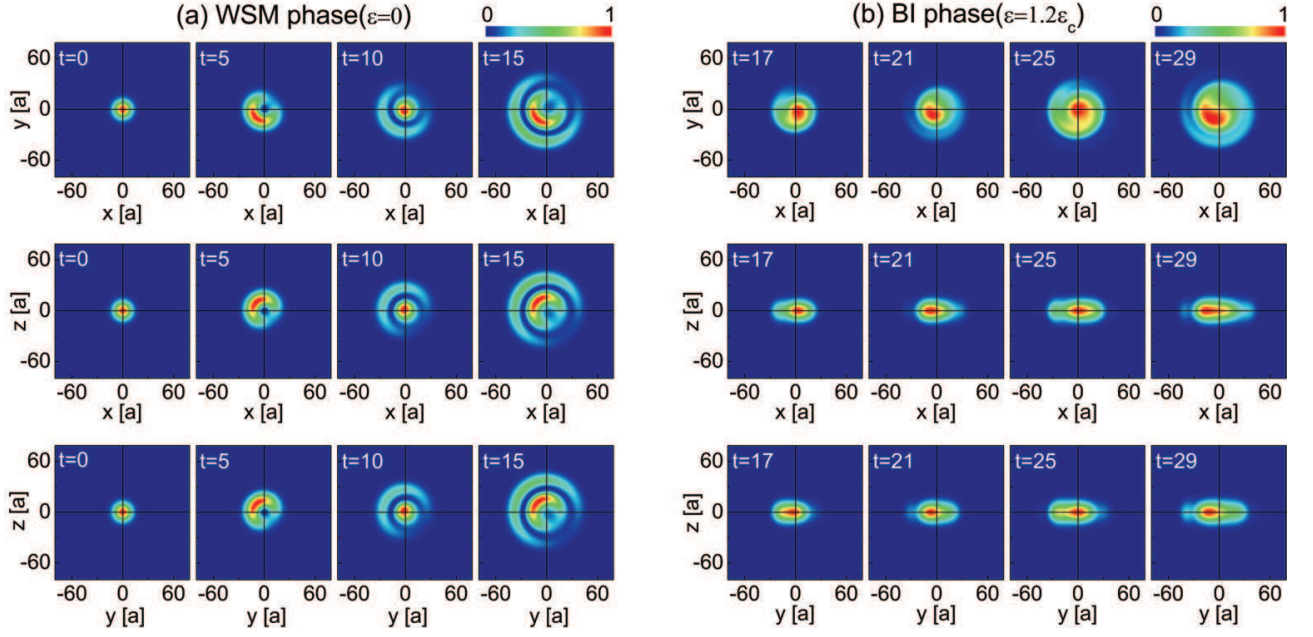}
 \label{Fig2} \caption{(Color online). The TOF snapshots
$|\Psi|^2$ with $L=10a$ at different $t$ (with the unit
$a/v_F=\hbar/(2J))$ in $x(y,z)=0$ cross-sections for (a) the WSM
phase with $\varepsilon=0$ and (b) the BI phase with
$\varepsilon=1.2\varepsilon_c$. The value of the probability is
rescaled from 0 to 1. }
\end{figure*}
%%%%%%%%%%%%%%%%%%%%%%%%%%%%%%%%%%%%%%%%%%%%%%%%%%%%%%%%%%%%%%%%%%%%%%%%%%%%%%%%%%%%%%%%%%%%%%%%%%%%%%%%%%%%%%%%%%%%%%%%%%%%%%
%%%%%%%%%%%%%%%%%%%%%%%%%%%%%%%%%%%%%%%%%%%%%%%%%%%%%%%%%%%%%%%%%%%%%%%%%%%%%%%%%%%%%%%%%%%%%%%%%%%%%%%%%%%%%%%%%%%%%%%%%%%%%%

\section{The evolution of wave-packet dynamics} Realization of
the optical Weyl lattice with ultracold atoms will open a new
frontier of research in Weyl physics, especially one can exploit
the dynamics of the particles described by the Weyl equation,
which might be hard to study in a condensed matter system. Here we
consider a Bose-Einstein condensation (BEC, or a cold atomic
ensemble) initially described by a 3D Gaussian wave packet
\begin{equation}\label{WF}
\begin{split}
|\Psi\rangle=(1/\sqrt{2\pi})^3\int d^3\mathbf{q}[
(L/\sqrt{\pi})^{\frac{3}{2}}e^{-\frac{1}{2}L^{2}\mathbf{q}^2}e^{i\mathbf{q}\mathbf{r}}]|\Phi\rangle,
\end{split}
\end{equation}
with the width $L$. The initial spinor state is chosen as
$|\Phi\rangle=(\frac{1}{\sqrt{3-{\sqrt{3}}}},\frac{1}{\sqrt{3+{\sqrt{3}}}}e^{i\frac{\pi}{4}})$,
in which $\langle\Phi|\sigma_j|\Phi\rangle=1/\sqrt{3}$ $(j=x,y,z)$. The
Gaussian wave packet (\ref{WF}) can be formed by applying an
additional 3D isotropic harmonic trap on the Weyl lattices, and
then move the harmonic potential with a velocity
$\hbar\textbf{W}_{\pm}/m_A$ with $m_A$ being the atomic mass to
place the BEC at the Weyl point $\textbf{W}_{\pm}$. In real space,
such wave packet is also a Gaussian shape and can be written as
$|\Psi(\textbf{r})\rangle=(1/\pi L^2)^{3/4} \exp (i
\textbf{W}_{\pm}\cdot\textbf{r})\exp(-r^2/2L^2)|\Phi\rangle$,
where $\textbf{r}$ is the atomic position related to the center of
the harmonic trap. One can show that the wave packet
$|\Psi(\textbf{r})\rangle$ in real space is related to
$|\Psi\rangle$ in Eq. (\ref{WF}) (note that $\textbf{q}$ in Eq.
(\ref{WF}) is the atom momentum related to the Weyl point
$\textbf{W}_{\pm}$) in momentum space with just a Fourier
transform. After releasing the trap, the atoms would start
expanding in the 3D Weyl lattice, and hence the dynamics of the
atoms (such as the snapshots shown in Fig. 2 and the PCM in Figs.
3 and 4 below), which is described by the Weyl equation with the
Hamiltonian determined by Eq. (\ref{g}), can be exploited with the
TOF experiments.

We calculate the time-dependent density profiles (which can be
directly measured by the TOF experiments) by solving the Weyl
equation with the standard split-operator method (see Appendix
VI). The TOF snapshots of the $x=0$, $y=0$ and $z=0$
cross-sections with a width of the wave packet $L=10a$ are shown
in Fig. 2(a) for the WSM phase and Fig. 2(b) for the BI phase.
Since $J$ can be tuned between $0.17 \sim 2.0$ kHz
\cite{Aidelsburger1,Aidelsburger2}, the time for $t=15$ is about
$0.6 \sim 7.1$ ms. In the WSM phase, the 3D Guassian wave packet
gradually evolves into a two-layer spherical shell structure. The
pattern of the density profiles is determined by the interference
of the positive and negative energy states. The
destructive-interference shell between two
constructive-interference spherical shell has a width of ten
lattice sites for $t=15$ and thus can be readily detected. The
most significant features of the evolution of Weyl quasiparticles
will be completely revealed when $t<15$. Since Weyl quasiparticles
are then moving away from the interference area of positive and
negative energy states, the shape of wave packets will remain
unchanged after time $t>15$. In the BI phase, the density profiles
exhibit the periodic oscillations in $x$- and $y$-axis direction,
meanwhile the PCM of the atomic gas is confined in $z=0$ plane.
Since the oscillation of the PCM will become more salient after
the wave function evolves for a longer time (here $t>17$), we
choose different period of time in Fig. 2 to reveal the features
of Weyl and BI phases.

The PCM is a crucial quantity in the study of the dynamics of the
atomic gas. It can be experimentally determined through TOF data,
and in our theoretical calculation it can be obtained by the
integral $\mathbf{\bar{r}}(t)=\int \mathbf{r}(t)|\Psi
(\mathbf{r}(t)) |^2d^3\mathbf{r}$ with $|\Psi (\mathbf{r}(t))|^2$
being derived by the split-operator method. Alternatively, the PCM
can be calculated analytically in the Heisenberg picture, where
the position operator is given by
\begin{equation}
\mathbf{\hat{r}}(t)=e^{i\hat{H}_{\textrm{eff}}t/\hbar}\mathbf{\hat{r}}(0)e^{-i\hat{H}_{\textrm{eff}}t/\hbar}.
\end{equation}

After inserting $\hat{H}_{\textrm{eff}}$ into the expression, we
obtain
\begin{equation}
\label{PCM G1}
%\begin{aligned}
\hat{\mathbf{r}}(t)=\hat{\mathbf{r}}(0)+\beta_{\mathbf{r}}t+\gamma_{\mathbf{r}}[\exp({2i\hat{H}_{\textrm{eff}}t/\hbar})-1],
%\end{aligned}
\end{equation}
where the coefficients
\begin{equation}
\begin{split}
&\beta_{x,y}=v_{x,y}^2\hat{H}^{-1}_{\textrm{eff}}\hat{q}_{x,y},\\
&\beta_{z}=\hat{H}^{-1}_{\textrm{eff}}(\Delta+\alpha_{z}v_z\hat{q}_{z}+\frac{\hat{q}_z^2}{2m^*})(\alpha_{z}v_z+\frac{\hat{q}_z}{m^*}),\\
&\gamma_{x,y}=\frac{\hbar}{2iE_{\mathbf{q}}^{2}}(\alpha_{x,y}v_{x,y}\hat{H}_{\textrm{eff}}\sigma_{y,x}-v_{x,y}^{2}\hat{q}_{x,y}),\\
&\gamma_{z}=\frac{\hbar}{2iE_{\mathbf{q}}^{2}}[\hat{H}_{\textrm{eff}}\sigma_{z}-(\Delta+\alpha_{z}v_{z}\hat{q}_{z}+\frac{\hat{q}_{z}^2}{2m^*})]
(\alpha_{z}v_z+\frac{\hat{q}_z}{m^*}).
\end{split}
\end{equation}

Hereafter $\hat{\mathbf{r}}$ and $\hat{\mathbf{q}}$ ($\mathbf{r}$
and $\mathbf{q}$) are the coordinate and momentum operators
(variables), respectively. By using the initial wave function in
Eq. (\ref{WF}), the expectation values $\bar{x}(t)$, $\bar{y}(t)$
and $\bar{z}(t)$ can be obtained as
\begin{eqnarray}
\label{PCM G2}
\bar{x},\bar{y}(t)=&&\frac{\alpha_{x,y}}{\sqrt{3}}\frac{L^3}{\pi^\frac{3}{2}}\iiint\limits_{-\infty}^{+\infty}\{A_{x,y}t+\eta_{x,y}B_{x,y}[\cos(2E_qt)\nonumber
\\&&-1]+C_{x,y}\sin(2E_qt)\} e^{-L^2\mathbf{q}^2}
d^3\mathbf{q}, \end{eqnarray}
\begin{equation}
\label{PCM_Z}
\bar{z}(t)=\frac{\alpha_{z}}{\sqrt{3}}\frac{L^3}{\pi^\frac{3}{2}}\iiint\limits_{-\infty}^{+\infty}[A_{z}t+C_{z}\sin(2E_qt)]
e^{-L^2\mathbf{q}^2}d^3\mathbf{q},
\end{equation}
where $\eta_x=-1$, $\eta_y=+1$, the other coefficients
\begin{equation}
\begin{split}
\left\{
  \begin{aligned}
&A_{x,y}=v_{x,y}^3q_{x,y}^2/E_{\mathbf{q}}^2\\
&A_{z}=\frac{\hbar}{\alpha_{z}E_{\mathbf{q}}^2}(\alpha_{z}v_z+\frac{q_z}{m^*})[(\Delta+\alpha_{z}v_zq_z+\frac{q_z^2}{2m^*})^2]\\
&B_{x,y}=\frac{\hbar}{2E_{\mathbf{q}}^2}(\Delta+q_z^2/2m^*)\\
&C_{x,y}=\frac{\hbar}{2E_{\mathbf{q}}^3}(v_{x,y}E_q^2-v_{x,y}^3q_{x,y}^2)\\
&C_{z}=\frac{\hbar}{2\alpha_{z}E_{\mathbf{q}}^3}(\alpha_{z}v_z+\frac{q_z}{m^*})[E_{\mathbf{q}}^2-(\Delta+\alpha_{z}v_zq_z+\frac{q_z^2}{2m^*})^2]
\end{aligned}.
   \right.
\end{split}
\end{equation}
Therefore, the PCM of the system can be straightforwardly obtained
from Eqs. (\ref{PCM G2},\ref{PCM_Z}).

\subsection{The PCM in the WSM phase}
When $\varepsilon\ll\varepsilon_{c}$, $\Delta=0$ and the terms
related to $m^*$ can be safely neglected, then the parameters
$B_{x,y}=0$, $A_{x,y,z}=v_{x,y,z}^{3}q_{x,y,z}^2/E_{\mathbf{q}}^2$
and
$C_{x,y,z}=\frac{\hbar}{2E_{\mathbf{q}}^3}(v_{x,y,z}E_{\mathbf{q}}^2-v_{x,y,z}^3q_{x,y,z}^2)$.
Therefore, the expressions of $\bar{x}$, $\bar{y}$ and $\bar{z}$
have the same form given by
\begin{equation}\label{PCM_WSM}
\bar{r}(t)=\frac{\alpha_{r}}{\sqrt{3}}\frac{L^3}{\pi^\frac{3}{2}}\iiint\limits_{-\infty}^{+\infty}\{A_{r}t+C_{r}\sin(2E_qt)\}
e^{-L^2\mathbf{q}^2}d^3\mathbf{q}.
\end{equation}

The motion of PCM can be divided into two parts: the first term is
a directed linear term with a constant anomalous velocity which
plays a crucial role in the celebrated anomalous and spin Hall
effects \cite{Xiao, Nagaosa}, and the second term is a short-lived
ZB term originating from the interference of positive and
negative-energy states
\cite{Vaishnav,LeBlanc,Qu,Li,Ghosh,Zawadzki,Romera,Demikhovskii}.
Furthermore, since the integrand in the expression is always
positive, the signs of $\bar{r}(t)$ are completely determined by
$\alpha_{x,y,z}$, therefore we have
$sgn(\bar{x}\bar{y}\bar{z})=sgn(\alpha_x\alpha_y\alpha_z)$, which
implies that the sign of the PCM can be used to determine the
chirality of each Weyl point.

%%%%%%%%%%%%%%%%%%%%%%%%%%%%%%%%%%%%%%%%%%%%%%%%%%%%%%%%%%%%%%%%%%%%%%%%%%%%%%%%%%%%%%%%%%%%%%%%%%%%%%%%%%%%%%%%%%%%%%%%%%%%%%
%%%%%%%%%%%%%%%%%%%%%%%%%%%%%%%%%%%%%%%%%%%%%%%%%%%%%%%%%%%%%%%%%%%%%%%%%%%%%%%%%%%%%%%%%%%%%%%%%%%%%%%%%%%%%%%%%%%%%%%%%%%%%%
\begin{figure}[htbp] \centering
 %Requires \usepackage{graphicx}
\includegraphics[width=8cm]{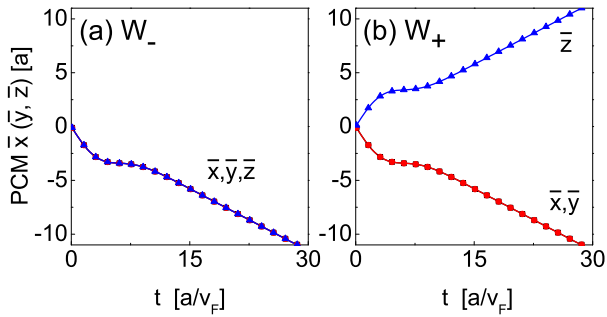}
\caption{(Color online). The expectation value $\bar{r}(t)$ for
the atomic gas initially at the $\mathbf{W}_{\pm}$ with
$\varepsilon=0$ and $L=10a$. The lines are obtained from Eq.
(\ref{PCM_WSM}), and the symbols are derived by the split-operator
method.}\label{Fig3}
\end{figure}
%%%%%%%%%%%%%%%%%%%%%%%%%%%%%%%%%%%%%%%%%%%%%%%%%%%%%%%%%%%%%%%%%%%%%%%%%%%%%%%%%%%%%%%%%%%%%%%%%%%%%%%%%%%%%%%%%%%%%%%%%%%%%%
%%%%%%%%%%%%%%%%%%%%%%%%%%%%%%%%%%%%%%%%%%%%%%%%%%%%%%%%%%%%%%%%%%%%%%%%%%%%%%%%%%%%%%%%%%%%%%%%%%%%%%%%%%%%%%%%%%%%%%%%%%%%%%

The Chern number of the valence band over a 2D sphere around a
Weyl point equals the Weyl point's chirality, which can be proved
as follows. The Berry curvature is given by
$\Omega_{ij}=\epsilon_{abc}g_{a}\partial_{i}g_{b}\partial_{j}g_{c}/(2E_{\mathbf{q}}^{3})$
\cite{Volovik}, and for the Hamiltonian (\ref{g}), we have
\begin{equation}
\label{BC}% Berry Curvature
\begin{split}
\mathbf{\Omega}=&-\frac{\alpha_{x}\alpha_{y}\hbar^{2}v_{x}v_{y}}{2E_{\mathbf{q}}^3}(\alpha_{z}v_{z}\hbar
q_{x}+\frac{\hbar^2q_{z}q_{x}}{m^*}, \\&\alpha_{z}v_{z}\hbar
q_{y}+\frac{\hbar^2q_{z}q_{y}}{m^*}, \Delta+\alpha_{z}v_{z}\hbar
q_{z}+\frac{\hbar^2q_{z}^2}{m^*}).
\end{split}
\end{equation}
As for WSM, the Berry curvature of each Weyl point can be derived
as $\mathbf{\Omega}_{\pm}=\mp
sgn(\alpha_{x}\alpha_{y}\alpha_{z})[\hbar^{3}v_{x}v_{y}v_{z}\mathbf{q}/2E_{\mathbf{q}}^3]$.
After integrating the whole sphere surrounding a Weyl point, we
derive the Chern number as
$N_C=sgn(\alpha_{x}\alpha_{y}\alpha_{z})$, and get the following
relation
\begin{equation}
\label{Ch}
N_C=N_\text{w}=sgn(\bar{x}\bar{y}\bar{z})=sgn(\alpha_{x}\alpha_{y}\alpha_{z}),
\end{equation}
which is $\pm 1$ for the Weyl points (see Appendix VI). The
expectation values $\bar{r}(t)$ for the atomic gas initially
located at the $\mathbf{W}_{\pm}$ with $\varepsilon=0$ and $L=10a$
are plotted in Fig. 3, where the relation (\ref{Ch}) is confirmed.
So the topological invariant of the system can be directly
detected from TOF data.

\subsection{The PCM in the BI phase}The system is in the BI
phase when $|\varepsilon|>\varepsilon_{c}$, in which pairwise Weyl
points have been merged and become the hybrid point featuring by a
hybrid spectrum shown in Fig. 1(e). The PCM $\bar{z}(t)=0$ because
the coefficients $A_{z}$ and $C_{z}$ in Eq. (\ref{PCM_Z}) are odd
function of $q_z$ when $v_{z}=0$. Then, the formula (\ref{Ch}) is
still valid and $N_{w}=sgn(\bar{x}\bar{y}\bar{z})=0$. Therefore,
the topological invariants of the system in both topological
trivial and non-trivial phases can be experimentally determined by
directly measuring the PCM.

%%%%%%%%%%%%%%%%%%%%%%%%%%%%%%%%%%%%%%%%%%%%%%%%%%%%%%%%%%%%%%%%%%%%%%%%%%%%%%%%%%%%%%%%%%%%%%%%%%%%%%%%%%%%%%%%%%%%%%%%%%%%%%
%%%%%%%%%%%%%%%%%%%%%%%%%%%%%%%%%%%%%%%%%%%%%%%%%%%%%%%%%%%%%%%%%%%%%%%%%%%%%%%%%%%%%%%%%%%%%%%%%%%%%%%%%%%%%%%%%%%%%%%%%%%%%%
\begin{figure}[htbp] \centering
 %Requires \usepackage{graphicx}
\includegraphics[width=8cm]{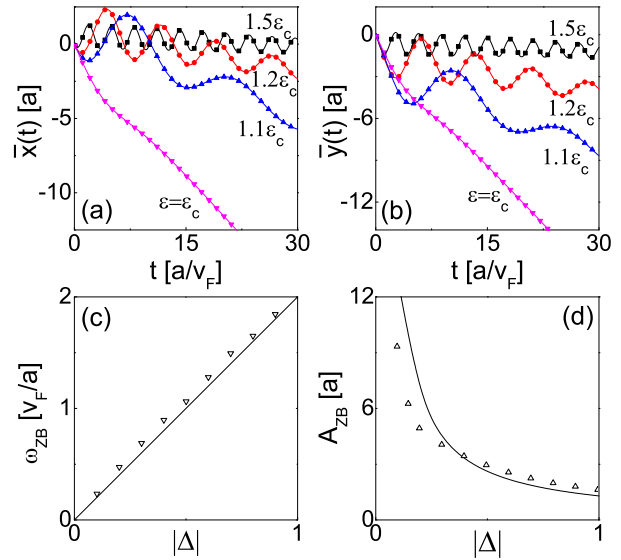}
\label{Fig4} \caption{(Color online). The expectation values (a)
$\bar{x}(t)$ and (b) $\bar{y}(t)$ in the BI phase with different
$\varepsilon$ and $L=10a$. The lines are obtained from Eq.
(\ref{PCM G2}) and the symbols are derived by the split-operator
method. (c) the frequency $\omega_{ZB}$ and (d) the amplitude
$A_{ZB}$ of \emph{Zitterbewegung} versus the energy gap $\Delta$.
The symbols are numerically calculated from Eq. (\ref{PCM G2}) and
the line of frequency (amplitude) is derived from the estimated
formula $\omega_{ZB}=2|\Delta|$ ($A_{ZB}=1.3/|\Delta|$).}
\end{figure}
%%%%%%%%%%%%%%%%%%%%%%%%%%%%%%%%%%%%%%%%%%%%%%%%%%%%%%%%%%%%%%%%%%%%%%%%%%%%%%%%%%%%%%%%%%%%%%%%%%%%%%%%%%%%%%%%%%%%%%%%%%%%%%
%%%%%%%%%%%%%%%%%%%%%%%%%%%%%%%%%%%%%%%%%%%%%%%%%%%%%%%%%%%%%%%%%%%%%%%%%%%%%%%%%%%%%%%%%%%%%%%%%%%%%%%%%%%%%%%%%%%%%%%%%%%%%%

To further characterize the dynamics of the PCM in the BI phase,
we plot the expectation values $\bar{x}(t)$ and $\bar{y}(t)$ with
different on-site energies in Fig. 4(a,b). Periodic oscillations
arise with the appearance of the gap, which is the direct evidence
of ZB effect. From Eq. (\ref{PCM G2}), we estimate the frequency
of ZB is $\omega_{ZB} \sim 2E_{\mathbf{q}}/\hbar \sim 2|\Delta|$
and the amplitude is $A_{ZB} \sim \sqrt{B^2_{x,y}+C^2_{x,y}} \sim
1.3/|\Delta|$. We also numerically calculate the $\omega_{ZB}$ and
$A_{ZB}$, and the results are plotted in Fig. 4(c,d), which agree
well with the estimations. At $\varepsilon=1.2\varepsilon_c$, the
amplitude $A_{ZB}$ is about three lattice sites and the period is
about $10 a/v_F$, which is around $0.4$ to $4.7$ ms for the
tunable $J$ between $0.17$ to $2.0$ kHz
\cite{Aidelsburger1,Aidelsburger2}. Such ZB oscillations can be
readily observed in TOF experiments.

\section{Conclusion}

%We now briefly address the experimental scheme to characterize the
%features in the Weyl adn BI phases through a visualized process of
%wave packet dynamics. Step 1, prepare Gaussian initial states
%satisfying Eq. (6) through a 3D harmonic trap. All initial states
%prepared are rest wave packets in real space. Step 2, move the
%prepared initial states into the Weyl lattice which is shown in
%Fig.1. Then by tuning parameters of the lattice, we make the
%objects to be tested (Weyl points or the hybrid point) located at
%the point q=0 in the momentum space. Meanwhile, make the center of
%initial states overlap the center of the objects to be observed in
%the momentum space. Step 3, remove the harmonic trap, let atoms
%evolve in the Weyl lattice. According to results of the density
%%distribution given by the manuscript, the system will be at WSM
%phase when the evolution is oriented, and BI phase if the
%evolution features an oscillation. Besides, the corresponding
%topological invariants can be determined by Equation (12). For
%instance, if the position of the center-of-mass (PCM) is detected
%in the first octant, the Weyl point to be observed has its Chern
%number $N_{c}=+1$. On the other hand£¬if PCM is detected in the
%second octant, then the Weyl point to be observed has its Chern
%number $N_{c}=-1$. "

In summary, we have exploited the dynamics of the Weyl
quasiparticles emerged in the optical lattices where the
topological WSM and trivial BI phases can be adjusted with the
on-site energy. We have demonstrated that the topological
invariants and the celebrated ZB effect can be directly observed
with TOF experiments. Since the dynamics of the Weyl particles
could be hard to detect in a condensed matter system, our proposal
in the atomic system would open up a novel possibility in research
of Weyl physics.

\section{acknowledgements}
We are grateful to Feng Mei and Xin Shen for useful discussions.
This work was supported by the NKRDP of China (Grant No.
2016YFA0301803), the NSFC (Grants No. 11474153 and 11604103), and
the PCSIRT (Grant No. IRT1243). D. W. Z. was supported by the NSF
of Guangdong Province (Grant No. 2016A030313436) and the Startup
Foundation of SCNU.

Z. L. and H. Q. W. contributed equally to this work.

\section{APPENDIX}
\subsection{The split-operator method}
It is generally known that the final wave function after an
evolution governed by the effective Hamiltonian [Eq. (\ref{g}) in
the main text] with time $t$, can be obtained as
\begin{equation}
\Psi(x,y,z,t)=\hat{\mathcal{T}}exp(-\frac{i}{\hbar}\int^{t}_{0}H_{\mathbf{eff}}(t')dt')\Psi(x,y,z,0),
\tag{A1}
\end{equation}
where $\hat{\mathcal{T}}$ denotes the the time ordering operator,
$\Psi(x,y,z,0)$ is the initial wave packet. The effective
Hamiltonian can be expressed as the sum of operators corresponding
to the kinetic and potential energies of the system in the form
$\hat{H_{\mathrm{eff}}}=\hat{T}+\hat{V}$. By using the standard
split-operator method, Eq.(A1) can be rewritten as
\begin{equation}
\begin{split}
\Psi(x,y,z,t+\delta
t)=&[e^{-\frac{i}{\hbar}\frac{\hat{T}}{2}\delta
t}e^{-\frac{i}{\hbar}\hat{V}\delta
t}e^{-\frac{i}{\hbar}\frac{\hat{T}}{2}\delta t}+\mathcal{O}(\delta
t^3)]
\\&\times\Psi(x,y,z,t),
\end{split}\tag{A2}
\end{equation}
where
$\hat{T}=\alpha_{x}v_{x}\hat{p}_{x}\sigma_{y}+\alpha_{y}v_{y}\hat{p}_{y}\sigma_{x}+(\alpha_{z}v_{z}\hat{p}_{z}+\frac{\hat{p}_{z}^2}{2m^*})\sigma_{z}$,
$\hat{V}=\Delta\sigma_{z}$. In the sufficiently short time $\delta
t$, the high-order term $\mathcal{O}(\delta t^3)$ (due to
noncommutation) can be safely neglected. One can connect the
position and the momentum spaces by using the Fourier transform.
Therefore, we can finally get the numerical solution of
$\Psi_{x,y,z,t}$ following the computation procedure step by step
with time step $\delta t$ \cite{Trotter}.

\subsection{The relations among the topological numbers and the
center-of-mass positions} There are eight Weyl points for the
Hamiltonian $H_k$ described in Eq. (\ref{Hk}) in the main text.
By requiring the coefficients of the Pauli matrices to be zero, we obtain
$W=[0$ or $\pi,
\pm\frac{\pi}{2},\pm\arccos(\frac{-\varepsilon}{2J_{z}})]$. By
expanding the quasi-momentum $\mathbf{k}$ at $W$ points,
$\mathbf{k}=W+\mathbf{q}$, we have
\begin{equation} \tag{B1}
\begin{split}
&\sin(k_{x}+q_{x})=\sin(k_{x})+\cos(k_{x})q_{x}-\frac{sin(k_{x})}{2}q_{x}^2
\\=&\left\{
  \begin{aligned}
+q_{x}, &\quad \mathrm{for}\quad k_{x}=0
\\
-q_{x}, &\quad \mathrm{for}\quad k_{x}=\pi
\end{aligned}.
   \right.
\end{split}
\end{equation}
\begin{equation} \tag{B2}
\begin{split}
&\cos(k_{y}+q_{y})=\cos(k_{y})-\sin(k_{y})q_{y}-\frac{\cos(k_{y})}{2}q_{y}^2
\\=&\left\{
  \begin{aligned}
-q_{y}, &\quad \mathrm{for}\quad k_{y}=\frac{\pi}{2}
\\
+q_{y}, &\quad \mathrm{for}\quad k_{y}=-\frac{\pi}{2}
\end{aligned}.
   \right.
\end{split}
\end{equation}
\begin{equation} \tag{B3}
\begin{split}
&\cos(k_{z}+q_{z})=\cos(k_{z})-\sin(k_{z})q_{z}-\frac{\cos(k_{z})}{2}q_{z}^2
\\=&\left\{
  \begin{aligned}
&-\frac{\varepsilon}{2J_{z}}-\sqrt{1-(\frac{-\varepsilon}{2J_{z}})^{2}}q_{z}+\frac{\varepsilon}{2J_{z}}\cdot\frac{q_{z}^2}{2},
 \\&\qquad\qquad\qquad\qquad \mathrm{for}\quad k_{z}=+arccos\frac{-\varepsilon}{2J_{z}}
\\
&-\frac{\varepsilon}{2J_{z}}+\sqrt{1-(\frac{-\varepsilon}{2J_{z}})^{2}}q_{z}+\frac{\varepsilon}{2J_{z}}\cdot\frac{q_{z}^2}{2},
\\&\qquad\qquad\qquad\qquad \mathrm{for}\quad k_{z}=-arccos\frac{-\varepsilon}{2J_{z}}
\end{aligned}.
   \right.
\end{split}
\end{equation}
Substituting them in Eq. (\ref{Hk}), we obtain the effective
Hamiltonian (\ref{g}). In the WSM phase, there exists a one-to-one
correspondence between the sign
($\alpha_{x},\alpha_{y},\alpha_{z}$) and the location of a Weyl
point ($k_{x},k_{y},k_{z}$). The relations are summed up in the
Table I.

%%%
\begin{table}[htbp]
\scalebox{1}[1]{%
\begin{tabular}{|c|c|c|c|c|}
\hline
\multicolumn{1}{|c|}{($k_{x},k_{y},k_{z}$)}  &\multicolumn{1}{|c|}{$(\alpha_{x},\alpha_{y},\alpha_{z})$} &\multicolumn{1}{|c|}{$N_w$} &\multicolumn{1}{|c|}{$N_C$} & octant\\
\hline
($0,+\frac{\pi}{2},+\arccos\frac{-\varepsilon}{2J_{z}}$)       &$ (-,+,+) $   &$ -1 $ &$ -1 $ & II  \\
\hline
($0,+\frac{\pi}{2},-\arccos\frac{-\varepsilon}{2J_{z}}$)      &$ (-,+,-) $   &$ +1 $ &$ +1 $ &VI  \\
\hline
($0,-\frac{\pi}{2},+\arccos\frac{-\varepsilon}{2J_{z}}$)     &$ (-,-,+) $   &$ +1 $ &$ +1 $ & III\\
\hline
($0,-\frac{\pi}{2},-\arccos\frac{-\varepsilon}{2J_{z}}$)      &$ (-,-,-) $   &$ -1 $ &$ -1 $ & VII\\
\hline
($\pi,+\frac{\pi}{2},+\arccos\frac{-\varepsilon}{2J_{z}}$)      &$ (+,+,+) $   &$ +1 $ &$ +1 $  & I \\
\hline
($\pi,+\frac{\pi}{2},-\arccos\frac{-\varepsilon}{2J_{z}}$)      &$ (+,+,-) $   &$ -1 $ &$ -1 $ &V \\
\hline
($\pi,-\frac{\pi}{2},+\arccos\frac{-\varepsilon}{2J_{z}}$)      &$ (+,-,+) $   &$ -1 $ &$ -1 $ & IV \\
\hline
($\pi,-\frac{\pi}{2},-\arccos\frac{-\varepsilon}{2J_{z}}$)      &$ (+,-,-) $   &$ +1 $ &$ +1 $ & VIII\\
\hline
\end{tabular}}
\caption{The relations among the locations of Weyl points, the
signs of Fermi velosity, winding number, Chern number and the
center-of-mass' position.}
\end{table}
%%%

In the BI phase $v_{z}=0$, the sign parameter $\alpha_{z}=0$. It
corresponds to the topological trivial state with winding number
$N_{w}=N_C=0$. Therefore, we obtain that
\begin{equation} \tag{B4}
\begin{split}
sgn(\bar{x}\bar{y}\bar{z})=sgn(\alpha_{x}\alpha_{y}\alpha_{z})=\left\{
  \begin{aligned}
\pm1, &\quad \text{for \ WSM}
\\
\quad 0, &\quad \text{for \ BI}
\end{aligned}.
   \right.
   \end{split}
\end{equation}

\subsection{Derivation of the Berry curvature} For the Hamiltonian
(\ref{g}) in the main text, the Berry curvature is given by
\cite{Volovik},
\begin{equation}\tag{C1}
\Omega_{ij}=\frac{1}{2g^{3}}\epsilon_{abc}g_{a}\partial_{i}g_{b}\partial_{j}g_{c}.
\end{equation}
It is straightforward to derive the following results
\begin{equation}\tag{C2}
\begin{split}
&\Omega_{xy}=\frac{1}{2g^{3}}\epsilon_{abc}g_{a}\partial_{x}g_{b}\partial_{y}g_{c}
=\frac{1}{2g^3}(-g_{3}\partial_{x}g_{2}\partial_{y}g_{1})
\\&=-(\Delta+\alpha_{z}v_{z}\hbar
q_{z}+\frac{\hbar^{2}q_{z}^{2}}{2m^{*}})\frac{\alpha_{x}\alpha_{y}\hbar^{2}v_{x}v_{y}}{2g^{3}}
\\&=\left\{
  \begin{aligned}
&-\frac{\alpha_{x}\alpha_{y}\alpha_{z}\hbar^{3}v_{x}v_{y}v_{z}}{2g^3}q_{z},
\quad\quad\quad \text{for \ WSM}
\\
&-(\Delta+\frac{\hbar^2q_{z}^2}{2m^{*}})\frac{\alpha_{x}\alpha_{y}\hbar^{2}v_{x}v_{y}}{2g^3},
\quad \text{for \ BI}
\end{aligned},
   \right.
\end{split}
\end{equation}
\begin{equation}\tag{C3}
\begin{split}
\Omega_{yz}&=\frac{1}{2g^{3}}\epsilon_{abc}g_{a}\partial_{y}g_{b}\partial_{z}g_{c}
=\frac{1}{2g^3}(-g_{2}\partial_{y}g_{1}\partial_{z}g_{3})
\\&=-q_{x}(\alpha_{z}v_{z}\hbar+\frac{\hbar^{2}q_{z}}{m^{*}})
\frac{\alpha_{x}\alpha_{y}\hbar^{2}v_{x}v_{y}}{2g^{3}}
\\&=\left\{
  \begin{aligned}
&-\frac{\alpha_{x}\alpha_{y}\alpha_{z}\hbar^{3}v_{x}v_{y}v_{z}}{2g^3}q_{x},
\quad\quad  \text{for \ WSM}
\\
&-(\frac{\hbar^{2}q_{z}q_{x}}{m^{*}})\frac{\alpha_{x}\alpha_{y}\hbar^{2}v_{x}v_{y}}{2g^3},
\quad \text{for \ BI}
\end{aligned},
   \right.
\end{split}
\end{equation}
\begin{equation}\tag{C4}
\begin{split}
\Omega_{zx}&=\frac{1}{2g^{3}}\epsilon_{abc}g_{a}\partial_{z}g_{b}\partial_{x}g_{c}
=\frac{1}{2g^3}(-g_{1}\partial_{z}g_{3}\partial_{x}g_{2})
\\&=-q_{y}(\alpha_{z}v_{z}\hbar+\frac{\hbar^{2}q_{z}}{m^{*}})
\frac{\alpha_{x}\alpha_{y}\hbar^{2}v_{x}v_{y}}{2g^{3}}
\\&=\left\{
  \begin{aligned}
&-\frac{\alpha_{x}\alpha_{y}\alpha_{z}\hbar^{3}v_{x}v_{y}v_{z}}{2g^3}q_{y},
\quad\quad \text{for \ WSM}
\\
&-(\frac{\hbar^{2}q_{z}q_{y}}{m^{*}})\frac{\alpha_{x}\alpha_{y}\hbar^{2}v_{x}v_{y}}{2g^3},
\quad \text{for \ BI}
\end{aligned}.
   \right.
\end{split}
\end{equation}
Then we have the general expression of Berry curvature in the
process of Weyl points' merging

\begin{equation}\tag{C5}
\begin{split}
\mathbf{\Omega}=&-\frac{\alpha_{x}\alpha_{y}\hbar^{2}v_{x}v_{y}}{2E_{\mathbf{q}}^3}(\alpha_{z}v_{z}\hbar
k_{x}+\frac{\hbar^2k_{z}k_{x}}{m^*},\alpha_{z}v_{z}\hbar
k_{y}\\&+\frac{\hbar^2k_{z}k_{y}}{m^*},\Delta+\alpha_{z}v_{z}\hbar
k_{z}+\frac{\hbar^2k_{z}^2}{m^*}).
\end{split}
\end{equation}
By tuning the parameter $\Delta$ [see Eq. (C5)], one can obtain
the corresponding Berry curvature in the whole process of
topological phase transition. We get, for the standard Weyl point,
\begin{equation}\tag{C6}
\mathbf{\Omega}=-sgn(\alpha_{x}\alpha_{y}\alpha_{z})\hbar^{3}v_{x}v_{y}v_{z}\mathbf{q}/(2E_{\mathbf{q}}^3),
\end{equation}
and for the hybrid point,
\begin{equation}\tag{C7}
\mathbf{\Omega}=-\frac{\alpha_{x}\alpha_{y}\hbar^{4}v_{x}v_{y}}{2m^*E_{\mathbf{q}}^3}(k_{x}k_{z},k_{y}k_{z},\frac{\Delta
m^*}{\hbar^2}+\frac{k_{z}^2}{2}).
\end{equation}
Thus, by integrating the whole sphere surrounding a Weyl/hybrid
point, one can easily obtain the corresponding Chern number.

%\end{CJK*}
\end{document}